\documentclass[twocolumn,pra,aps]{revtex4-1}
\usepackage[latin1]{inputenc}
\usepackage{amsmath}
\usepackage{amsfonts}
\usepackage{amssymb}
\usepackage{graphicx}
\usepackage{textcomp}
\begin{document}
\title{ A Davidson-Lanczos iteration method for computation of continued-fraction expansion of the Green's function at very low 
temperatures: Applications to the dynamical mean field theory}
\author{Medha Sharma}
\email{medhajamia@gmail.com}
\author{M.A.H. Ahsan}
\email{mahsan@jmi.ac.in}
\affiliation{Department of Physics, Jamia Millia Islamia, New Delhi 110025, India}

\begin{abstract}
We present a combination method based on orignal version of Davidson algorithm for extracting few of the lowest eigenvalues and eigenvectors
of a sparse symmetric Hamiltonian matrix and the simplest version of Lanczos technique for obtaining a tridiagonal representation of the Hamiltonian to compute the continued fraction expansion of the Green's function at a very low temperature. We compare the Davidson$+$Lanczos method with the full diagonalization on a one-band Hubbard model on a Bethe lattice of infinite-coordination using dynamical mean field theory.
\end{abstract}
\maketitle
\section{Introduction}

The infinite lattice
coordination limit introduced by Metzner and Vollhardt\cite{mv1989}
forms the basis for the Dynamical Mean Field Theory(DMFT)\cite{ag1992}
that maps the Hubbard model\cite{jh1963} onto an Anderson impurity model(AIM)\cite{pw1961}.
Even though the spatial degrees of freedom are completely frozen and the Anderson impurity model is
much simpler than the  original lattice model, it is still a nontrivial many-body problem.
In practice, the most difficult step in the DMFT iterative procedure is the repeated calculation of the impurity Green's function $G(i\omega_{n})$ of the impurity problem for any given arbitrary conduction electron effective bath $(G_{0}(i\omega_{n}))$.
Anderson impurity model can be solved either by numerical methods like exact diagonalization(ED)\cite{ck1994}, quantum Monte Carlo(QMC)\cite{hf1986},
numerical renormalization group(NRG)\cite{rb1999} or by analytic methods like iterated perturbation theory(IPT)\cite{ag1992}. 
Most of these methods have limitations confining them to a particular regime, ie high temperature(QMC) or low temperature(ED, NRG).

Exact Diagonalization(ED)\cite{ed1994} is an important technique for studying quantum many-body systems.
Green's function at finite temperature can be computed using all the eigenvalues and eigenvectors obtained via full diagonalization. 
Full diagonalization needs an explicit representation of the matrix, requiring lot of memory space. Therefore it is limited to small clusters because of large memory required for an exponentially growing Hilbert space.
Moreover, for diagonalizing an $n\times n$ matrix, $O(n^{3})$ floating point operations(flops) are required. But if the matrix is sparse and
a few eigenvalues and eigenvectors are required then we can resort to iteration methods like the Lanczos\cite{ln1950} method or the Davidson\cite{ed1975}
method. Green's function at zero temperature can computed by continued fraction expansion using Lanczos coefficients.
Green's function at very low temperature can be computed by the set of equations used by Capone et al\cite{mc2007}.

Lanczos method is an implementation of the Rayleigh Ritz procedure\cite{ys2011} on a Krylov subspace\cite{ak1931} whereas 
Davidson method is on non-Krylov subspace.
In this paper, we use the orignal version of Davidson method and the simplest version of the Lanczos algorithm and find that
the inclusion of Davidson method in evaluation of Green's function in low temperature regime can be favourably used to our advantage.

The contents of this paper are organised as follows. In section II., we describe the orthogonal projection method explaining the Lanczos and davidson algorithm. In section III., we discuss the computation of Green's function at different temperature regimes.
Section IV. gives a sketch of DMFT procedure. In Section V., we show the comparision between the Davidson$+$Lanczos method of computation of Green's function and the full ED. Finally in section VI., we discuss the advantages of this combination method.

\section{Orthogonal projection method}

Consider the eigenvalue problem: ${\bf{H}}u = \lambda u$
where $\bf{H}$ is an $n\times n$ matrix, $u$ $\in $ $R^{n}$ and $\lambda $ $\in $ $R$ .
An orthogonal projection method \cite{ys2011} finds an approximate eigenvalue $\tilde{\lambda} \in R $ and eigenvector
$\tilde{u}$ $\in $ $S$(say an s-dimensional subspace of $R^{n}$)
which satisfy the Galerkin condition, thereby making
the residual vector of $\tilde{u}$ orthogonal to the subspace $S$, i.e.,
$ {\bf{H}}\tilde{u}-\tilde{\lambda}\tilde{u} \bot S $.
Therefore, $\left\langle {\bf{H}}\tilde{u}-\tilde{\lambda}\tilde{u}|q\right\rangle=0$ , $\forall q \in S$, where
${q_{1} , q_{2} , . . . , q_{s}}$ is an orthogonal basis of $S$ and
$Q $ is a matrix with column vectors $q_{1} , q_{2} , . . . , q_{s}$.
Let $\tilde{u} = Qy$, yielding $\left\langle {\bf{H}}Qy-\tilde{\lambda}Qy|q_{j}\right\rangle=0$, where $j = 1,2,...,s$.
This leads to the relation $P_{s}y =\tilde{\lambda}y$, where $P_{s} = Q^{T}{\bf{H}}Q$, so $y$ and $\tilde{\lambda}$ must be the
eigenvalue and eigenvector of the matrix $P_{s}$ respectively.

\subsection{Rayleigh-Ritz Procedure}
The numerical procedure for computing the Galerkin approximations to solve the eigenvalue problem
is known as the Rayleigh-Ritz procedure. We describe the procedure in a step-by-step way as follows:
\begin{enumerate}
\item Compute an orthonormal basis ${q_{i}}$ of a $s$ dimensional subspace $S$. Form an $n \times s $ 
   matrix whose columns are $q_{1},q_{2},...q_{s}$, where $s<<n$.
\item Compute the $ s \times s $ matrix $P_{s} = Q^{T}{\bf{H}}Q$.
\item Compute the desired eigenvalue $\tilde{\lambda_{k}}$ and eigenvector $y_{k}$ of the small matrix $P_{s}$, 
      where $k<=s$.
\item Compute the corresponding eigenvector of ${\bf{H}}$, $\tilde{u_{k}}=Qy_{k}$.
\end{enumerate}

\subsubsection{Davidson Method}

Davidson\cite{ed1975} proposed an iterative calculation for the diagonalization of large, sparse, real-symmetric
matrices to find a few of lowest eigenvalues and corresponding eigenvectors.
Davidson method generates an orthonormal set of $s$ basis vectors onto which a projection 
of the matrix to be diagonalized is performed and a correction vector is computed if the residual vector of the 
computed eigenvector is not a null vector.

Let $\tilde{\lambda_{k}}$ be desired the Ritz eigenvalue and $\tilde{u_{k}}$ be the Ritz eigenvector 
of the eigenvalue problem ${\bf{H}}u=\lambda u$ and $k<=s$. Let $c$ be an n-component correction vector.
\begin{eqnarray*}
     u_{k}=\tilde{u_{k}}+c\\
     {\bf{H}}(\tilde{u_{k}}+c)=\lambda_{k}(\tilde{u_{k}}+c)\\
     (\lambda_{k}I-{\bf{H}})c=({\bf{H}}-\lambda_{k}I)\tilde{u_{k}}\\
     \lambda_{k} \approx \tilde{\lambda_{k}}\\
     (\tilde{\lambda_{k}}I-{\bf{H}})c=({\bf{H}}\tilde{u_{k}}-\tilde{\lambda_{k}}\tilde{u_{k}})\\
     c=\frac{({\bf{H}}\tilde{u_{k}}-\tilde{\lambda_{k}}\tilde{u_{k}})}{(\tilde{\lambda_{k}} I-{\bf{H}})}\\
     c=\frac{({\bf{H}}\tilde{u_{k}}-\tilde{\lambda_{k}}\tilde{u_{k}})}{(\tilde{\lambda_{k}} I-P)}
\end{eqnarray*}
where, P is is some preconditioning matrix. In the original algorithm, P was simply the main diagonal of the matrix ${\bf{H}}$.

This correction vector is made orthogonal to all the previous vectors and then added to the basis set. This procedure
is repeated until convergence, i.e. a null residual vector. If the dimension of the subspace spanned by these basis vectors becomes inconveniently
large due to successive augmentation of the subspace, the calculation is restarted with the first $k$ Ritz eigenvectors. If several eigenvectors 
are sought, then the first $s$ Ritz eigenvectors obtained at the end of finding one eigenvector provides a good starting set for the next eigenvector.

The Jacobi-Davidson\cite{sv1996} method of solving the sparse eigenvalue problem combines the Davidson\cite{ed1975} method and the 
Jacobi's\cite{jr1846} approach and has improved convergence properties.
But we use the orignal Davidson method taking the preconditioning matrix as the diagonal elements of the Hamiltonian matrix to compute its groundstate and a few excited states and give reasons for our choice in Section VI..
 
We find $nval$ eigenvalues and eigenvectors of $n\times n$ Hamiltonian matrix ${\bf{H}}$ using Davidson method in Algorithm I, given in Appendix \ref{algo1},
where $diag$ is a one-dimensional array of length $dm $(dimension of ${\bf{H}}$) storing all the diagonal elements of the ${\bf{H}}$ and we start with $ndv$ number of trial vectors which is greater than $nval$. We avoid the
storage of the vectors obtained after the matrix vector multiplication ($hvec$) which is usually done in the implementation of the Davidson
algorithm.
In Appendix B, we discuss a practical aspect of coding of Davidson algorithm.

\subsubsection{Lanczos Method}

Lanczos\cite{ln1950} method is used for diagonalizing large sparse hermitian matrices for computation of extremal(smallest or largest) eigenvalues. 
The basic idea of Lanczos method is the construction of basis where the projection of the Hamiltonian has a tridiagonal representation.
The optimization\cite{gl1996} of Rayleigh quotient $r(x)=x^{T}{\bf{H}}x/x^{T}x$, where ${\bf{H}}\in R^{n\times n}$  and  $x\neq0\in R^{n} $, leads to the problem of computing orthonormal basis ${q_{1},...q_{n}}$ for the Krylov subspace\cite{ak1931} $ {K_{j}({\bf{H}},q_{1})=q_{1},{\bf{H}}q_{1},...,{\bf{H}}^{j-1}q_{1}}$.

We choose an arbitrary starting unit norm vector $q_{1}$ having a non-zero overlap with the actual groundstate vector and define the successive vector as ${\bf{H}}q_{1}$ and make it orthogonal against $q_{1}$ using the Gram-Schmidt procedure to obtain the second unit norm vector $q_{2}$,
\begin{eqnarray*}
q^\prime_{2} = {\bf{H}} q_{1} - (q^{T}_{1}{\bf{H}} q_{1})q_{1}\\
\left|\left|q^\prime_{2}\right|\right|q_{2} = {\bf{H}}q_{1} - (q^{T}_{1}{\bf{H}}q_{1})q_{1},
\end{eqnarray*}
where the symbol $\left|\left|. \right|\right|$ denotes a vector 2-norm.
Similarly the third unit norm vector $q_{3}$ is obtained by defining it as ${\bf{H}}q_{2}$ and making it orthogonal against $q_{2}$ and $q_{1}$,
\begin{eqnarray*}
q^\prime_{3} = {\bf{H}}q_{2} - (q^{T}_{2}{\bf{H}}q_{2})q_{2}-(q^{T}_{1}{\bf{H}}q_{2})q_{1}\\
\left|\left|q^\prime_{3}\right|\right|q_{3} = {\bf{H}}q_{2}-(q^{T}_{2}{\bf{H}}q_{2})q_{2}-(q^{T}_{1}{\bf{H}}q_{2})q_{1}.
\end{eqnarray*} 
Thus the Gram-Schmidt orthogonalization of the basis of Krylov subspace can be generalized to the following reccurence relation,
as we have to orthogonalize the vector only to the previous two vectors, and the orthogonality with the earlier ones is automatic,
atleast in exact arithmetic:

\begin{equation}
{\bf{H}}q_{n}=\alpha(n) q_{n}+\beta(n)q_{n-1}+\beta(n+1)q_{n+1}, 
\end{equation}
where $\alpha(n)=q_{n}^{T}{\bf{H}}q_{n}$, $\beta(n)=q_{n-1}^{T}{\bf{H}}q_{n}=q_{n}^{T}{\bf{H}}q_{n-1}$.

In this basis, the Hamiltonian matrix reduces to a symmetric tridiagonal matrix:
\begin{eqnarray}
\left[
\begin{array}{lllll}
\alpha(1) & \beta(2)  & 0         & 0         & \cdots\\ 
\beta(2)  & \alpha(2) & \beta(3)  & 0         & \cdots\\
0         & \beta(3)  & \alpha(3) & \beta(4)  & \cdots\\
0         & 0         & \beta(4)  & \alpha(4) & \cdots\\
\vdots    & \vdots    & \vdots    & \vdots    & \ddots\\
\end{array}
\right]
\end{eqnarray} 
We construct $\alpha^{,}$s and $\beta^{,}$s in Algorithm II, given in Appendix \ref{algo2}, where $x$ is an arbitrary starting vector of length $ dm $(dimension of the {\textbf{H}}) and $itermax$ is the maximum number of planned lanczos steps, typically about 100(determined by preliminary tests). The array $\alpha(1:nlan)$ forms the diagonal and the array $\beta(2:nlan)$ forms the sub-daigonal of the tridiagonal matrix, $nlan $ being the actual number of iterations carried out.

The simplest version of Lanczos method without any form of reorthogonalization has been used in the evaluation of Green's function as it turns out to be sufficient to obtain the tridiagonal representation of {\textbf{H}}.

\section{Green's function}
We use the exact diagonalization technique to find the eigenvalues and eigenvectors of ${\bf{H}}$.
We perform diagonalization in a sector$(N_{\uparrow},N_{\downarrow})$ of total Hilbert space with
fixed number of spin up electrons and of spin down electrons. We compute the Green's function by using the basis states of
one spin configuration $(\sigma=\uparrow or \downarrow)$ at a time\cite{ms2014}.

\subsection{Green's function at finite temperature}

At finite temperature, the Green's function is computed using the expression:
 
\begin{equation}
G_{\sigma}(p,i\omega_{n})=1/Z\sum_{i,j}
\frac{\left|\left\langle i|c^{\dagger}_{p,\sigma}|j\right\rangle\right|^{2}}{E_{j}-E_{i}+i\omega_{n}}(e^{-\beta E_{i}}+e^{-\beta E_{j}}),
\end{equation}

where $Z$ is the partition function and $c_{p \sigma}^{\dagger}$ is fermion creation operator at site $p$ with spin $\sigma (\uparrow or \downarrow)$ and
$\omega_{n}$ is the matsubara frequency. The full set of states $\left|i\right\rangle(\left|j\right\rangle)$ are the eigenvectors with corresponding eigenvalue $E_{i}(E_{j})$. The full set of eigenvalues $E_{j}(E_{i})$ are of the sector $(N_{\sigma},N_{\overline{\sigma}})((N_{\sigma}+1,N_{\overline{\sigma}}))$, where if $\sigma=\uparrow(\downarrow)$ then $\overline{\sigma}=\downarrow(\uparrow)$ and $N_{\sigma}$ varies from $0$ to $M-1$ and $N_{\overline{\sigma}}$ varies from $0$ to $M$,
wherein $M$ is total number of sites.

All the eigenvalues and eigenvectors are obtained by full diagonalization which is feasible upto $M=8$, where the largest sector
$(N_{\uparrow}=4,N_{\downarrow}=4)$ has dimension 4900.

\subsection{Green's function at zero temperature}

We evaluate the zero temperature Green's function using continued fraction expansion\cite{hh1972,pf1991,ed1994}.

The zero temperature Green's function is expressed as:
\begin{equation*}
G_{\sigma}(p,i\omega_{n})=G_{\sigma,e}(p,i\omega_{n})+G_{\sigma,h}(p,i\omega_{n})=
\end{equation*}
\begin{equation}
\left\langle\psi_{0}\right|c_{p,\sigma}\frac{1}{i\omega_{n}+(E_{0}-{\bf{H}})}c^{\dagger}_{p,\sigma}\left|\psi_{0}\right\rangle+\nonumber
\end{equation}
\begin{equation}
\left\langle\psi_{0}\right|c^{\dagger}_{p,\sigma}\frac{1}{i\omega_{n}-(E_{0}-{\bf{H}})}c_{p,\sigma}\left|\psi_{0}\right\rangle
\end{equation}
where $G_{\sigma,e}(p,i\omega_{n})(G_{\sigma,h}(p,i\omega_{n}))$ describe the electron(hole) excitation, ${\bf{H}}$ is the Hamiltonian matrix, $E_{0}$ and $\psi_{0}$ are the groundstate eigenvalue and eigenvector of ${\bf{H}}$ respectively.

First we find the groundstate eigenvalue and eigenvector using the Lanczos method by starting with a arbitrary unit vector.
Then we express ${\bf{H}}$ in tridiagonal form by applying the Lanczos method but instead starting the iteration with an arbitrary unit vector, we use
$ q_{1}=\phi_{1e}
=\frac{c^{\dagger}_{p,\sigma}\left|\psi_{0}\right\rangle}{\sqrt{\left\langle\psi_{0}\right|c_{p,\sigma}c^{\dagger}_{p,\sigma}\left|\psi_{0}\right\rangle}}
(\phi_{1h}=\frac{c_{p,\sigma}\left|\psi_{0}\right\rangle}{\surd{\left\langle\psi_{0}\right|c^{\dagger}_{p,\sigma}c_{p,\sigma}\left|\psi_{0}\right\rangle}})$ 
as the starting unit norm vector and compute the orthonormal set of basis vectors $q_{1},q_{2},...q_{m}$ for evaluation of $G_{\sigma,e}(p,i\omega_{n})(G_{\sigma,h}(p,i\omega_{n}))$.

Consider the matrix $(z-esign{\bf{H}})$ and the identity $(z-esign{\bf{H}})(z-esign{\bf{H}})^{-1}=I$, where $z=i\omega_{n}+esign(E_{0})$, $esign=+1(-1)$ for electron(hole)
excitation. Expressing in the above Lanczos basis starting with $\phi_{1e}(\phi_{1h})$, we obtain $\sum_{n}(z-esign{\bf{H}})_{gn}(z-esign{\bf{H}})_{nr}^{-1}=\delta_{gr}$.
For the special case $r=1$, we get $\sum_{n}(z-esign{\bf{H}})_{gn}y_{n}=\delta_{g1}$, where $y_{n}=(z-esign{\bf{H}})^{-1}_{n1}$. This represents a system of equations:

\begin{eqnarray}
(z-esign{\bf{H}})_{11}y_{1}+\cdots(z-esign{\bf{H}})_{1n}y_{n}=1\nonumber\\
(z-esign{\bf{H}})_{21}y_{1}+\cdots(z-esign{\bf{H}})_{2n}y_{n}=0\nonumber\\
\vdots\nonumber\\
(z-esign{\bf{H}})_{n1}y_{1}+\cdots(z-esign{\bf{H}})_{nn}y_{n}=0
\end{eqnarray}

We want to evaluate the quantity, $y_{1}=\left\langle\phi_{1f}\right|\frac{1}{z-esign{\bf{H}}}\left|\phi_{1f}\right\rangle$, where $f$ denotes $e(h)$ for electron(hole). Therefore, by Cramer's rule we get
\begin{equation}
y_{1}=\frac{detD_{esign}^{\prime}(2)}{detD_{esign}(1)},
\end{equation} 
where the matrices are expressed in ${\left|\phi_{1f}\right\rangle,\left|\phi_{2f}\right\rangle,\cdots \left|\phi_{nf}\right\rangle}$, i.e., ${q_{1},q_{2},\cdots q_{n}}$.
  
\begin{eqnarray*}
D_{esign}(1)=z-esign({\bf{H}})=
\end{eqnarray*}
\begin{eqnarray}
\left[
\begin{array}{llll}
z+esign (-\alpha(1)) & esign(-\beta(2))    & \cdots\\
esign (-\beta(2))    & z+esign(-\alpha(2)) & \cdots\\
0                    & esign(-\beta(3))    & \cdots\\
0                    & 0                   & \cdots\\
\vdots               & \vdots              & \ddots\\
\end{array}
\right],
\end{eqnarray}
where $\alpha(1:n)$ and $\beta(2:n)$ are the coefficients calculated in the Lanczos procedure.\\

\begin{eqnarray*}
D^{\prime}_{esign}(2)=
\end{eqnarray*}
\begin{eqnarray}
\left[
\begin{array}{llll}
1                    & esign(-\beta(2))    & \cdots\\
0                    & z+esign(-\alpha(2)) & \cdots\\
0                    & esign(-\beta(3))    & \cdots\\
0                    & 0                   & \cdots\\
\vdots               & \vdots              & \ddots\\
\end{array}
\right]
\end{eqnarray}

\begin{eqnarray*}
D_{esign}(2)=
\end{eqnarray*}
\begin{eqnarray}
\left[
\begin{array}{lll}
z+esign(-\alpha(2)) & esign(-\beta(3))    & \cdots\\
esign(-\beta(3))    & z+esign(-\alpha(3)) & \cdots\\
0                   & esign(-\beta(4))    & \cdots\\
\vdots              & \vdots              & \ddots\\
\end{array}
\right]
\end{eqnarray}

From above we see that $DetD(2)^{\prime}=DetD_{esign}(2)$.
Matrix $D_{esign}(n+1)$ is obtained from the matrix $D_{esign}(1)$ by removing the first $n$ rows and columns.
By using the Laplace expansion of the first row of $DetD_{esign}(1)$, we obtain $DetD_{esign}(1)=[z-esign(\alpha(1))]DetD_{esign}(2)-\beta(2)^{2}DetD_{esign}(3)$ which can be generalized to give a recurrence relation: 
\begin{equation*}
DetD_{esign}(n)=[z-esign(\alpha(n))]DetD_{esign}(n+1)
\end{equation*}
\begin{equation}
-\beta(n+1)^{2}DetD_{esign}(n+2).
\end{equation}
Therefore,
\begin{equation}
y_{1}= \frac{1}{[z-esign(\alpha(1))]-\beta(2)^{2}\frac{DetD_{esign}(3)}{DetD_{esign}(2)}},
\end{equation}
which can be further written as
\begin{equation*}
y_{1}= 
\end{equation*}
\begin{equation}
\frac{1}{[z-esign(\alpha(1))]-\beta(2)^{2}\frac{1}{[z-esign(\alpha(2))]-\beta(3)^{2}\frac{DetD_{esign}(4)}{DetD_{esign}(3)}}},
\end{equation}
giving rise to a continued fraction expansion.\\

Finally, Green's function can be evaluated as:
\begin{equation*}
 G_{\sigma}(p,i\omega_{n})=G_{\sigma,e}(p,i\omega_{n})+G_{\sigma,h}(p,i\omega_{n})=
\end{equation*}
\begin{equation}
\frac{||c^{\dagger}_{p,\sigma}\left|\psi_{0}\right\rangle||^{2}}{DetD_{+1}(1)/DetD_{+1}(2)}+\frac{||c_{p,\sigma}\left|\psi_{0}\right\rangle||^{2}}{DetD_{-1}(1)/DetD_{-1}(2)}.
\end{equation}

\subsection{Green's function at very low temperature}

To compute the Green's function at very low temperature we give the set of equations used by Capone et al\cite{mc2007}.

We start with the spectral representation of Green's function is given as:

\begin{equation}
G_{\sigma}(p,i\omega_{n})=1/Z\sum_{i,j}\frac{\left|\left\langle i| c^{\dagger}_{p,\sigma}|j\right\rangle\right|^{2}}{E_{j}-E_{i}+i\omega_{n}}(e^{-\beta E_{i}}+e^{-\beta E_{j}}),
\end{equation}

This is written in compressed form as:

\begin{equation}
G_{\sigma}(p,i\omega_{n})=1/Z\sum_{i}e^{-\beta E_{i}} G_{\sigma}^{i}(p,i\omega_{n}) \label{count5},
\end{equation}
where $Z=\sum_{i}e^{-\beta E_{i}}$

\begin{equation}
G_{\sigma}^{i}(p,i\omega_{n})=\sum_{j}\frac{\left|\left\langle j| c_{p,\sigma}|i\right\rangle\right|^{2}}{E_{j}-E_{i}+i\omega_{n}}+
\sum_{j}\frac{\left|\left\langle j| c^{\dagger}_{p,\sigma}|i\right\rangle\right|^{2}}{E_{i}-E_{j}+i\omega_{n}},
\end{equation}

The Green's function $G_{\sigma}^{i}(p,i\omega_{n})$ is computed using continued fraction expansion of Lanczos coefficients
obtained by taking the initial trial vectors as $c^{\dagger}_{p,\sigma}\left|i\right\rangle$ for particle excitation and $c_{p,\sigma}\left|i\right\rangle$ for hole excitation respectively.
At low temperatures, the Boltzmann factors in (\ref{count5}) ensures that only a small number of excited states are required.
The computation of these excited states has been done using the Lanczos algorithm\cite{mc2007} which is plagued by the loss of orthogonality
in the Lanczos basis leading to incorrect degeneracy. The computation of these excited states has also been done by using the Arnoldi
method \cite{gl1996} of which the Lanczos is a special case for a Hermitian matrix\cite{ca2007}.
In our method, we use the orignal version of the Davidson method\cite{ed1975} to compute these excited states.
 
\section{DMFT procedure}
The DMFT procedure\cite{ag1996} is given as follows:

An initial guess of the Anderson parameters $[\varepsilon_{l},V_{l}]$, that defines the  Anderson impurity model:
\begin{eqnarray} 
H_{Anderson}=\varepsilon_{d}\sum_{\sigma} d_{\sigma}^{\dagger}d_{\sigma}+\sum_{l=2,\sigma}^{m}\varepsilon_{l}a_{l\sigma}^{\dagger}a_{l\sigma}+\nonumber\\
Un_{d\uparrow}n_{d\downarrow}+\sum_{l=2,\sigma}^{m}V_{l}(a_{l\sigma}^{\dagger}d_{\sigma}+d_{\sigma}^{\dagger}a_{l,\sigma})\label{count6}
\end{eqnarray}
($d^{\dagger}_{\sigma}$ and $a^{\dagger}_{l,\sigma}$ are creation operators for fermions
associated with the impurity site and with the state $l$ of the effective bath, respectively) is made.
The $ U=0 $ Green's function of the impurity(we drop the spin index of Green's function):
\begin{equation}
G_{0}(i\omega_{n})^{-1}=i\omega_{n}+\mu-\sum_{l=1}^{Ns}\frac{|V_{l}|^{2}}{i\omega_{n}-\varepsilon_{l}}
\end{equation}
is computed.
After the Anderson impurity model is solved by Exact Diagonalization, the impurity Green's function:
\begin{equation}
G(i\omega_{n})=\int_{0}^{\beta}d\tau e^{i\omega_{n}\tau}(-\left\langle T_{ \tau}c(\tau)c^{\dagger}(0)\right\rangle)
\end{equation} 
is calculated and the self energy is extracted using the Dyson equation:
\begin{equation}
\Sigma_{imp}(i\omega_{n})=G_{0}^{-1}(i\omega_{n})-G^{-1}(i\omega_{n})
\end{equation} 
The self energy becomes local in the infinite coordination limit:
\begin{equation}
\Sigma(i\omega_{n})\approx\Sigma_{imp}(i\omega_{n})
\end{equation}
The onsite Green's function:
\begin{equation}
G(i\omega_{n})=\int d\varepsilon \frac{D(\varepsilon)}{i\omega_{n}+\mu-\varepsilon-\Sigma(i\omega_{n})}
\end{equation} 
which depends on the noninteracting density of states $D(\varepsilon)$ of the orignal lattice is computed.

The bath Green's function is  updated by using the Dyson equation again
\begin{equation}
G_{0}(i\omega_{n})^{new}=[G^{-1}(i\omega_{n})+\Sigma(i\omega_{n})]^{-1}.
\end{equation}
A cost function of the form (in the present work, we have taken the following cost function),
\begin{equation}
\chi=\frac{1}{nmax+1}\sum_{0}^{nmax}|G_{0}(i\omega_{n})^{new}-G_{0}(i\omega_{n})|
\end{equation}

is minimized to obtain a new set of parameters $[\varepsilon_{l},V_{l}]$, where $nmax$ is very large upper cut off and 
$|.|$ is the square root of sum of the
squares of the differences of the real and imaginary parts of $G_{0}(i\omega_{n})^{new}$ and $G_{0}(i\omega_{n})$. We use the
mimimization conjugate gradient routine minimize provided in Ref. 14 to find these parameters.

The above process is repeated till convergence.

For an infinite coordination Bethe lattice with semicircular density of states of half bandwidth $D$, the self consistency equation
reduces to:
\begin{equation}
G_{0}(i\omega_{n})^{-1}=i\omega_{n}+\mu-\frac{D^{2}}{4}G(i\omega_{n})
\end{equation}
and for a paramagnetic normal state, the self consistency equation becomes:
\begin{equation}
G_{0}(i\omega_{n})^{-1}=i\omega_{n}+\mu-G(i\omega_{n})/2.\label{count7}
\end{equation}
For half filled case, $\mu=\frac{U}{2}$. For convience we have taken $\varepsilon_{d}=-\frac{U}{2}$ and made use of shifted chemical
potential $\varDelta\mu=\mu-\frac{U}{2}=0$.

\section{Results}

We show results for the paramagnetic half-filled Hubbard model using the self consistency eq.(\ref{count7}).

\begin{figure}[h]
\resizebox{8.5cm}{!}{\includegraphics[width=4.5cm]{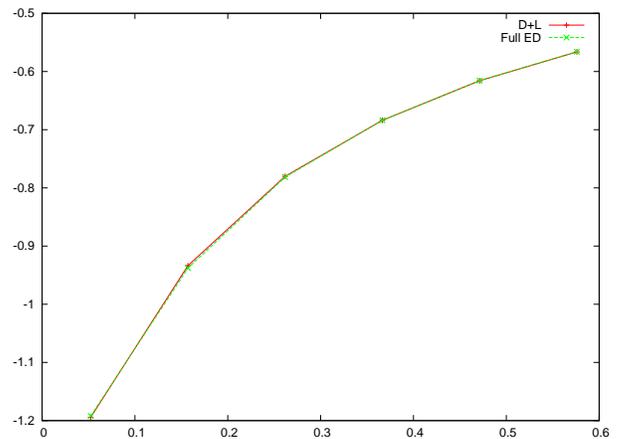}} 
\caption{Imaginary part of the local Green's function on the matsubara axis using Davidson$+$Lanczos method compared with the full diagonalization for $M=6$ and $\beta=60$ with $U=3$, $\epsilon_{d}=-1.5$}
\end{figure}
\begin{figure}[h]
\resizebox{8.5cm}{!}{\includegraphics[width=4.5cm]{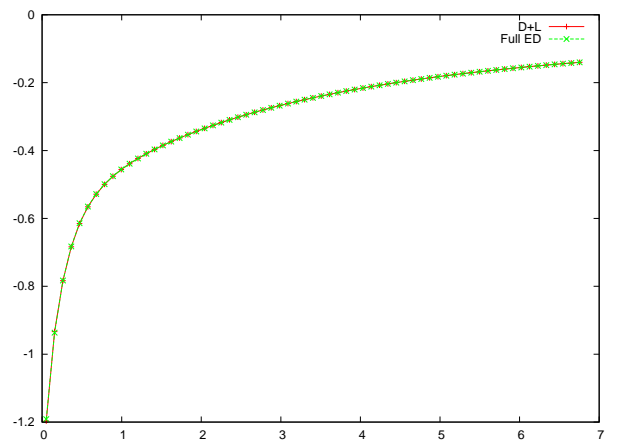}} 
\caption{Imaginary part of the local Green's function on the matsubara axis using Davidson$+$Lanczos method compared with the full diagonalization for $M=8$ and $\beta=60$ with $U=3$, $\epsilon_{d}=-1.5$}
\end{figure}

Fig.1. and Fig.2. show that the Green's function obtained with Davidson+Lanczos method is almost exactly the same as obtained using full diagonalization for $M=6$ and $M=8$ with the inclusion of $6$ and $20$ eigenstates respectively.

\section{Conclusions}
If we analyse the sparse Hamiltonian matrix of the Anderson impurity model, eqn.(\ref{count6}), we find the following observations about 
the {\textbf{diagonal part}} of {\textbf{H}} that are favourable for the application of the Davidson method:\\

$\bullet$ Each spin up or spin down electron residing on any bath site $l$(impurity site $d$) will add a term $\epsilon_{l}(\epsilon_{d})$
and if the impurity site is doubly occupied then an additional term $U$ is added  to the diagonal part of {\textbf{H}}. All finite $\epsilon_{l}$'s make sure that even a single spin up or spin down electron in any sector $(N_{\uparrow},N_{\downarrow})$ makes all diagonal elements of the sparse matrix {\textbf{H}} non-zero (except the one in which the electron is on the impurity site and the diagonal element is given by the sum of $\epsilon_{d}$ and $U$). Thus {\textbf{H}} is {\textbf{diagonally dominant}}\cite{tj1980}.

$\bullet$ For an Anderson impurity model on an $M$ site lattice in which the transition is
possible between the impurity site and the bath constituted by all other
sites, each row of {\textbf{H}} will have a maximum of $2M-2$ nonzero off-diagonal elements. The total number of off-diagonal non-zero elements of {\textbf{H}} of a sectror $(N_{\uparrow},N_{\downarrow})$ is
$(2M-2)\times{^{M-2}C}_{N_{\uparrow}-1}\times{^{M}C}_{N_{\downarrow}}+(2M-2)\times{^{M-2}C}_{N_{\downarrow}-1}\times{^{M}C}_{N_{\uparrow}} $, where $M>2$, $0<N_{\uparrow}<M$ and $0<N_{\downarrow}<M$. Except for sectors $(M,M)$ or $(0,0)$ where the dimension of {\textbf{H}} is 1, in all other sectors, {\textbf{H}} is {\textbf{never completely diagonal}}(for finite $V_{l}$'s), thereby not making the trial vectors linearly dependent.
 
$\bullet$ The contribution of $U$ to the diagonal part of {\textbf{H}} when the impurity site is doubly occupied and not the same exact numerical values of all $\epsilon_{l}$'s and $\epsilon_{d}$ further confirm that the {\textbf{diagonal}} of {\textbf{H}} is {\textbf{not}} a {\textbf{constant}}, thereby not making Davidson method equivalent to Lanczos method.\\\\

In the present work, we use the simplest version of both the Lanczos and Davidson method without going into 
any of their advanced variants(band or block versions)\cite{gl1996} that makes them loose their simplicity to compute the Green's function 
at a very low temperature. The inclusion of Davidson method to calculate the excited states helps us in the following ways:\\

$\bullet$ We rule out the Ghost eigenvalues\cite{cw1979} which are replicas of converged
eigenvalues and are the consequence of the rounding off errors that leads to the loss of orthogonality among the Lanczos vectors
as after a few iterations they start containing large components of the dominant eigenvectors as in the case of power method.

$\bullet$ We correctly resolve the degeneracy of the system and properly determine the multiplicity.

$\bullet$ Convergence is better and faster, i.e. more accurate eigenpairs are obtained in lesser number of iterations\cite{ed1975}.

\begin{acknowledgements}
Medha Sharma is thankful to DST for financial assistance in the form
of Inspire Fellowship. Medha Sharma also thanks The Institute of Mathematical Sciences, Chennai, India for a visit.
\end{acknowledgements}

\appendix

\section{Algorithm I}
\label{algo1}
\noindent $\lbrace$ Choose \textbf{$ ndv $} arbitrary guess vectors each of dimension \textbf{$ dm $} and make them orthonormal with respect to each other using Modified Gram-Schmidt procedure and store them in a two-dimensional array \textbf{$ vec $}($1:dm $,$ 1:ndv $).$\rbrace$\\
$n1=ndv-1$\\
\textbf{while} $ival<=nval$\\
\indent\indent $ps(1:n1,1:n1)=0$\\
\indent\indent\textbf{for} $k=1:n1 $\\
\indent\indent\indent\indent\textbf{for} $ j=1:dm $\\
\indent\indent\indent\indent\indent\indent $hvec(j)= H(j,:)vec(:,k)$\\
\indent\indent\indent\indent\textbf{end}\\
\indent\indent\indent\indent\textbf{for} $ i=1:k $\\
\indent\indent\indent\indent\indent\indent\textbf{for} $ j=1:dm $\\
\indent\indent\indent\indent\indent\indent\indent\indent $ps(i,k)=ps(i,k)+vec(j,i)hvec(j)$\\
\indent\indent\indent\indent\indent\indent\textbf{end}\\
\indent\indent\indent\indent\textbf{end}\\
\indent\indent\textbf{end}\\
\indent\indent\textbf{for} $j=1:n1-1 $\\
\indent\indent\indent\indent\textbf{for} $ i=j+1:n1 $\\
\indent\indent\indent\indent\indent\indent $ ps(i,j)=ps(j,i) $\\
\indent\indent\indent\indent\textbf{end}\\
\indent\indent\textbf{end}\\
\indent\indent $\lbrace $ Compute all eigenvalues $ wr(1:*)$ and all eigenvectors $ zr(1:n1,1:*)$ of the $ n1 \times n1 $ matrix \textbf{$ps$}
using QR algorithm (standard subroutine). $\rbrace $\\
\indent\indent\textbf{for} $ i=1:dm $\\
\indent\indent\indent\indent $ o(1:n1)=0 $\\
\indent\indent\indent\indent\textbf{for} $ j=1:n1 $\\
\indent\indent\indent\indent\indent\indent\textbf{for} $ k=1:n1 $\\
\indent\indent\indent\indent\indent\indent\indent\indent $ o(j)=o(j)+vec(i,k)zr(k,j)$\\
\indent\indent\indent\indent\indent\indent\textbf{end}\\
\indent\indent\indent\indent\textbf{end}\\
\indent\indent\indent\indent\textbf{for} $ j=1:n1 $ \\
\indent\indent\indent\indent\indent\indent $ vec(i,j)=o(j) $\\
\indent\indent\indent\indent\textbf{end}\\
\indent\indent\textbf{end}\\
\indent\indent $ cnvgd=true $\\
\indent\indent\textbf{for} $ i=1:dm $\\
\indent\indent\indent\indent $hvec(i)= H(i,:)vec(:,ival)$\\
\indent\indent\textbf{end}\\
\indent\indent\textbf{if} $n1>=ndv$\\
\indent\indent\indent\indent\textbf{for} $ i= 1:dm $ \\
\indent\indent\indent\indent\indent\indent $res=hvec(i)- wr(ival)vec(i,ival)$\\
\indent\indent\indent\indent\indent\indent\textbf{if} $ abs(res)> 0 $\\
\indent\indent\indent\indent\indent\indent\indent\indent $ cnvgd=false $\\
\indent\indent\indent\indent\indent\indent\textbf{end}\\
\indent\indent\indent\indent\textbf{end}\\
\indent\indent\textbf{else}\\
\indent\indent\indent\indent\textbf{for} $i=1:dm$\\
\indent\indent\indent\indent\indent $vec(i,n1+1)=hvec(i)- wr(ival)vec(i,ival)$\\
\indent\indent\indent\indent\indent\indent\textbf{if} $ abs(vec(i,n1+1))> 0 $\\
\indent\indent\indent\indent\indent\indent\indent\indent $ cnvgd=false $\\
\indent\indent\indent\indent\indent\indent\textbf{end}\\
\indent\indent\indent\indent\textbf{end}\\
\indent\indent\textbf{end}\\
\indent\indent\textbf{if} $ cnvgd=false $ and $n1<ndv $\\
\indent\indent\indent\textbf{for} $ i=1:dm $\\
\indent\indent\indent\indent\textbf{if} $ abs(wr(ival)-diag(i))>0 $\\
\indent\indent\indent\indent\hspace{.07 in}$ vec(i,n1+1)=vec(i,n1+1)/wr(ival)-diag(i)$\\
\indent\indent\indent\indent\textbf{end}\\
\indent\indent\indent\textbf{end}\\
\indent\indent\textbf{end}\\
\indent\indent\textbf{if} $cnvgd=true$\\
\indent\indent\indent\indent $ival=ival+1 $\\
\indent\indent\textbf{else}\\
\indent\indent\indent\indent$n1=n1+1$\\
\indent\indent\textbf{end}\\
\indent\indent\textbf{if} $ n1<ival $ or $ n1=ndv+1 $\\
\indent\indent\indent\indent $n1=ival$\\
\indent\indent\textbf{end}\\
\indent\indent $\lbrace$ Construct the first \textbf{$ n1 $} vectors each of dimension \textbf{$ dm $} stored in two-dimensional array {\textbf{$ vec $}($1:dm $,$ 1:n1 $)} orthonormal with respect to each other using Modified Gram-Schmidt procedure. $\rbrace$\\
\textbf{end}\\

\section{An important practical aspect while coding of Davidson algorithm to find a few extremal eigenvalues and eigenvectors}
When we are soughting several eigenvalues, we proceed to find the next eigenvalue once the just previous eigenvalue has converged, i.e., its
residual norm is zero. This implies that all the converged eigenvalues are the eigenvalues of the large orignal matrix or in other words, they all are Ritz values. But they may not be the correct sequential (increasing or decreasing) eigenvalues of the orignal matrix. It is possible that
some of the eigenvalues have been missed in between.

When a few eigenvalues have converged, the projection matrix obtained at that point usually modifies the earlier eigenvalues. We generally 
encounter the following cases:
$\bullet$ The eigenvalues obtained when they are converged one by one, say, in increasing order are exactly the same (small difference $\varepsilon$ of the order of $10^{-13}$) as the eigenvalues of the projection matrix obtained after several eigenvalues have converged and these
eigenvalues are the correct {\bf{sequential}} eigenvalues of the orignal large matrix.\\
$\bullet$ The Ritz values are obtained when the eigenvalues are converged one by one in increasing order but they are not the correct 
{\bf{sequential}} eigenvalues of the orignal large matrix which are given by the eigenvalues of the projection matrix obtained after the convergence of several eigenvalues.\\
$\bullet$The Ritz values are obtained when the eigenvalues are converged one by one in increasing order but they are not the correct {\bf{sequential}} eigenvalues and the eigenvalues of the projection matrix obtained after convergence of several  eigenvalues are in the process of becoming the correct sequential eigenvalues and at that point may be incorrect eigenvalues, i.e., not even Ritz values.\\
$\bullet$The eigenvalues obtained when they are converged one by one in increasing order are exactly the same (small difference $\varepsilon$ of the order of $10^{-13}$) as the eigenvalues of the projection matrix obtained after several eigenvalues have converged but they are {\bf{not}}
the correct {\bf{sequential}} eigenvalues of the orignal large matrix although being Ritz values.\\

To deal with all the above cases, the number of trial vectors that forms the basis of the subspace, i.e., the dimension $(ndv)$ of the subspace in which the orignal matrix is to be projected is chosen to be sufficiently large and each eigenvalue from $1$ to a large number (maximum upto $ndv-1$) is converged one by one to obtain the projection matrix whose first few eigenvalues are most likely to be the correct {\bf{sequential}} eigenvalues of the orignal large matrix.

\section{Algorithm II}
\label{algo2}
\noindent $ nlan=0 $\\
\textbf{for} $ iter=1:itermax $\\
\indent\indent$ nlan=nlan+1 $\\
\indent\indent\textbf{if} $iter=1$\\
\indent\indent\indent\indent\textbf{for} $ i=1:dm $\\
\indent\indent\indent\indent\indent\indent $ qold(i)=x(i)/\left|\left|x\right|\right| $\\
\indent\indent\indent\indent\textbf{end}\\
\indent\indent\indent\indent $ qnew(1:dm)=0 $\\
\indent\indent\textbf{end}\\
\indent\indent\textbf{if} $iter \ne 1$\\
\indent\indent\indent\indent \textbf{for} $ i=1:dm $\\
\indent\indent\indent\indent\indent\indent $ qvold=qold(i)$\\
\indent\indent\indent\indent\indent\indent $ qold(i)=qnew(i)/\beta(iter) $\\
\indent\indent\indent\indent\indent\indent $ qnew(i)=-\beta(iter)qvold $\\
\indent\indent\indent\indent \textbf{end}\\
\indent\indent\textbf{end}\\
\indent\indent\textbf{for} $ i=1:dm $\\
\indent\indent\indent\indent $ hqold(i)=H(i,:)qold(:)$\\
\indent\indent\textbf{end}\\
\indent\indent $ qnew(:)=qnew(:)+hqold(:) $\\
\indent\indent $ \alpha(iter)=dot_{\_}product(qold,qnew) $ \\
\indent\indent $ qnew(:)=qnew(:)-\alpha(iter)qold(:) $\\
\indent\indent $ \beta(iter+1)=\left|\left|qnew\right|\right| $\\
\indent\indent \textbf{if} $\beta(iter+1)=0 $\\
\indent\indent\indent\indent\textbf{break}\\
\indent\indent\textbf{end}\\
\textbf{end}

\end{document}